\documentclass[twocolumn,showpacs,preprintnumbers,amsmath,amssymb,pra]{revtex4}
\usepackage{color,latexsym,epsfig,amssymb,amsfonts,amsmath,graphicx,bbm,fancyhdr,enumerate}

\newcommand{\ket}[1]{\left | \, #1 \right \rangle}
\newcommand{\bra}[1]{\left \langle #1 \, \right |}

\newcommand{\outprod}[2]{\ket{#1}\bra{#2}}

\begin{document}

\title{Entanglement fidelity of quantum memories}

\author{K.~Surmacz, J.~Nunn, F.~C.~Waldermann, Z.~Wang, I.~Walmsley, and D.~Jaksch}

\affiliation{Clarendon Laboratory, University of Oxford, Parks
Road, Oxford OX1 3PU, United Kingdom}

\begin{abstract}
We introduce a figure of merit for a quantum memory which measures
the preservation of entanglement between a qubit stored in and
retrieved from the memory and an auxiliary qubit. We consider a
general quantum memory system consisting of a medium of two level
absorbers, with the qubit to be stored encoded in a single photon.
We derive an analytic expression for our figure of merit taking
into account Gaussian fluctuations in the Hamiltonian parameters,
which, for example, model inhomogeneous broadening and storage
time dephasing. Finally we specialize to the case of an atomic
quantum memory where fluctuations arise predominantly from Doppler
broadening and motional dephasing.
\end{abstract}
\date{\today}

\maketitle

\begin{figure}
\begin{center}
\includegraphics[width=8.5cm]{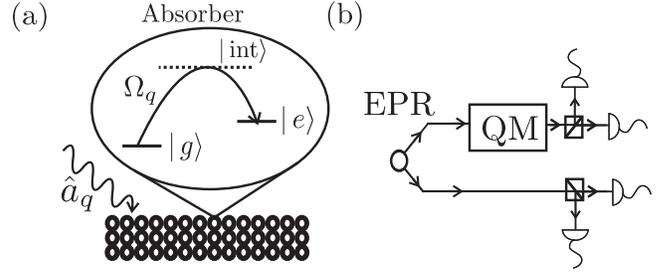}
\end{center}
\caption{(a) General level structure of an absorber in a QM. A
photon with annihilation operator $\hat{a}_q$ is incident on a
medium of $N$ absorbers, and excites one absorber into state
$\ket{e}$ via an intermediate state $\ket{\mathrm{int}}$. (b)
Schematic experimental setup. We consider a photonic qubit
entangled with an auxiliary qubit produced by an EPR source. The
photonic qubit is stored in the memory, and the amount of
entanglement that remains after storage is measured.}
\label{Fig_setup}
\end{figure}

The ability to store flying qubits in a quantum memory (QM) is a
fundamental component of many quantum communication
schemes~\cite{briegeletal,duretal}. Numerous possible methods for
storing and retrieving qubits encoded in light pulses have been
proposed~\cite{fleischhauerlukin1,polzik,kraus}, and some of these
proposals have recently been experimentally realized, achieving
e.g.~storage and retrieval of a single photon on
demand~\cite{chaneliere,julsgaard}, and entanglement between light
and matter~\cite{monroe,polzik}. Many promising candidate systems
for QMs such as atomic ensembles \cite{lukin}, arrays of quantum
dots \cite{santori} or NV centers in diamond \cite{kurtsiefer} can
often effectively be described as ensembles of $N$ two-level
absorbers coupling to the incoming qubit. We consider two
independent ensembles each storing one of the logical qubit
states. The absorbers consist of two meta-stable internal states
$\ket{g}$ and $\ket{e}$ as shown in Fig.~\ref{Fig_setup}(a), and a
transition $\ket{g}\rightarrow\ket{e}$ is effected by the incoming
photon in logical state $q$ via coupling $\Omega_q$. The states
$\ket g$ and $\ket e$ are usually not directly connected
optically, with this transition often being achieved via an
intermediate state $\ket{\mathrm{int}}$ and additional control
fields. For most of this paper details of such additional
structure in the absorbing medium are not considered, and we
assume that its effects on the properties of the absorbers can be
subsumed into stochastic fluctuations of the coupling parameter
$\Omega_q$. After a storage time $t_s$ another control field is
used to retrieve the photonic qubit. Dephasing may take place in
the memory during the storage time, which usually leads to
different couplings when writing and reading the qubit.

Using these general assumptions and the notion of entanglement
fidelity \cite{schumacher} we derive a figure of merit $\cal F$
that measures how well a QM setup can preserve entanglement
between a qubit undergoing the memory process (the memory qubit)
and an auxiliary qubit. Our figure of merit $\cal F$ is different
from commonly-used quality measures such as average fidelity $F_A$
for a pre-defined set of input qubit states~\cite{hammerer}. This
captures the ability of a memory to recreate the initial state of
the qubit, and is equal to $1$ if and only if the memory stores
and retrieves every state perfectly. However, depending on the
application of the QM, one might not necessarily be concerned with
exactly preserving the quantum state of the qubit. The
preservation of entanglement might be more desirable in some
quantum information processing and quantum communication schemes
\cite{ekert1,bennett}, for example in a quantum repeater
\cite{briegeletal,duretal} or in the cascaded generation of graph
states \cite{raussendorf}. The entanglement fidelity $\cal F$ also
directly relates to the degree of violation of a Bell inequality
by an EPR pair of photons, where one photon is stored and
subsequently retrieved from the QM while the auxiliary qubit is
directly detected as schematically shown in
Fig.~\ref{Fig_setup}(b). The setup shown in
Fig.~\ref{Fig_setup}(b) could thus be used to measure our figure
of merit.

In the system outlined above the memory qubit is encoded in a
subspace of the overall photon Hilbert space $\mathcal{H}_A$. The
states of the memory and auxiliary qubits are denoted by $\ket{0}$
and $\ket{1}$. The Hilbert spaces of the auxiliary qubit and the
medium are $\mathcal{H}_B$ and $\mathcal{H}_C$ respectively. The
system has initial state
$\hat{\rho}_0=\outprod{\phi_0}{\phi_0}\otimes\hat{\rho}_C$, where
$\ket{\phi_0}\in\mathcal{H}_A\otimes\mathcal{H}_B$ and
$\hat{\rho}_C$ is the initial density operator of the medium. We
assume that the absorbers are not correlated initially and
characterize a QM as a quantum operation $\Lambda_M$ that acts on
the photon as follows
\begin{equation}
\label{charmem}
\Lambda_M\otimes\mathbb{I}:\outprod{\phi_0}{\phi_0}\rightarrow\mathrm{tr}_C\left[\hat{\mathcal{L}}(\hat{\rho}_0)\right]\mbox{,}
\end{equation}
where $\hat{\mathcal{L}}$ is a Liouvillian operating on states in
$\mathcal{H}_A\otimes\mathcal{H}_C$, and $\mathbb{I}$ is the
identity operator on states in $\mathcal{H}_B$. We note that in
the work of A.~K.~Ekert et al.~\cite{ekert} a quantum channel for
qubits $\Lambda$ is characterized by considering the action of the
operator $\Lambda\otimes\mathbb{I}$ on two qubit states. The
superoperator $\Lambda_M\otimes\mathbb{I}$ preserves entanglement
for all two-qubit states only if $\Lambda_M$ is unitary (the
converse is well known \cite{plenio}). This can be seen by using a
Kraus decomposition of $\Lambda_M$. We find that for at least one
initial two-qubit state $\ket{\phi_e}$ the application of a
non-unitary $\Lambda_M$ will result in a mixed state. Purification
of $(\Lambda_M\otimes\mathbb{I})(\outprod{\phi_e}{\phi_e})$
results in the introduction of an extra ancillary system, with
which the memory qubit is entangled. By monogamy of
entanglement~\cite{wootters,koashi,bruss}, the entanglement
between the memory and auxiliary qubits decreases.

Motivated by these observations we write a QM entanglement
fidelity as follows
\begin{equation}
\label{fid1}
\mathcal{F}(\Lambda_M)=\min_{\ket{\phi_0}}\left\{\bra{\phi_0}\hat{U}_{M}^{\dagger}\left[(\Lambda_M\otimes\mathbb{I})(\outprod{\phi_0}{\phi_0})\right]\hat{U}_{M}\ket{\phi_0}\right\}\mbox{.}
\end{equation}
The quantity inside the braces is the entanglement fidelity
\cite{schumacher} for the process $\Lambda_M$ applied to the state
$\mathrm{tr}_B[\outprod{\phi_0}{\phi_0}]$ ($\mathrm{tr}_B$ denotes
the partial trace over $\mathcal{H}_B$). The entanglement fidelity
was introduced as a measure to characterize how well entanglement
is preserved by such a process in \cite{schumacher}, and detailed
discussions of its properties can be found in
\cite{schumacher,nielsen,kretschmann}. Since the standard
definition of entanglement fidelity \cite{nielsenchuang} measures
preservation of state as well as entanglement we include a unitary
$\hat{U}_M$, which acts on $\mathcal{H}_A$, to allow for evolution
of the photon that would not decrease the entanglement present.
This unitary is chosen to maximize $\cal F$, and thus describes an
optimized storage process to which $\Lambda_M$ is compared in the
same way that gate fidelity \cite{nielsenchuang} measures the
success of a quantum gate. We also minimize over all pure
two-qubit input states so that $\mathcal{F}$ is a property only of
the QM that uses the worst-case scenario as a measure of its
success. The QM $\Lambda_M$ (and hence the Liouvillian
$\hat{\mathcal{L}}$) consists of a read-in process, a period of
storage, and a read-out process that retrieves the photon on
demand a time $t_s$ after read-in. Note that more sophisticated
choices for $\hat{U}_M$ conditional on the outcome of measurements
on the state of the QM after retrieving the photon might enable
further improvement of $\cal F$. However, such schemes are
difficult to realize experimentally and are not considered in this
paper. Thus if $\mathcal{F}=1$ we have that $\Lambda_M$ preserves
entanglement between the qubits, but the final and initial states
of the photon may be deterministically different. The
representation of the QM with $\Lambda_M$ illustrates that for
${\cal F}<1$ the memory process will not be unitary.

We now consider the photon and its interaction with the ensemble of
absorbers. We define the annihilation operator $\hat{a}_q$ for the
photon in state $\ket{\underline q}=\hat{a}_q^{\dagger}
\ket{\mathrm{vac}}$, where $\ket{\mathrm{vac}}$ represents the
vacuum state and $q=0,1$ denotes the logical state of the qubit (the
underline distinguishes states in $\mathcal{H}_A$ from memory qubit
states). This annihilation operator can be written as
\begin{equation}
\label{annih}
\hat{a}_q=\int\mathrm{d}\mathbf{k}g_q(\mathbf{k})\hat{a}_{\mathbf{k},\lambda_q}\mbox{,}
\end{equation}
where $\hat{a}_{\mathbf{k},\lambda_q}$ destroys a photon with
polarization $\lambda_q$ and wavevector $\mathbf{k}$. The mode
functions $g_q(\mathbf{k})$ are normalized,
$[\hat{a}_q,\hat{a}_{q'}^{\dagger}]=\delta_{qq'}$ and for
simplicity we have assumed that each logical state has an
associated single polarization $\lambda_q$. The absorbers are
initially in the collective state $\ket{G}=\ket{g_1,\dots,g_N}$,
and are assumed to coherently couple to the photon during the
whole of the read-in and read-out processes. The Hamiltonian for
the read-in interaction between the photon in state
$\ket{\underline{q}}$ and the $j^{\mathrm{th}}$ absorber is given
by $
\hat{H}_q^{(j)}=(\Omega_{q,j}\hat{a}_{q,j}\hat{\sigma}_{eg}^{(j)}+\mathrm{H.c.})$,
where $\hat{\sigma}^{(j)}_{eg}=\ket{e}_j\bra{g}$. During storage
each absorber evolves according to the Hamiltonian
$\hat{H}_{S,q}^{(j)}=s_{q,j}(t)\hat{\sigma}_{ee}^{(j)}$, with
$s_{q,j}(t)$ some time-dependent detuning. The read-out
interaction of the photon in logical state $\ket{q}$ with the
absorber is modeled by the Hamiltonian $
\hat{\tilde{H}}_q^{(j)}=(\tilde{\Omega}_{q,j}\hat{b}_{q,j}\hat{\sigma}_{eg}^{(j)}+\mathrm{H.c.})$,
with couplings $\tilde{\Omega}_{q,j}$. The dependence of the
operators $\hat{a}_{q,j}$ and $\hat{b}_{q,j}$ on the absorber
reflects the fact that due to motion each absorber will in general
couple to a slightly different mode. We assume that an appropriate
choice of control field can restrict this effect to a phase
$\delta_{q,j}^{(a)}$, so that $\hat{a}_{q,j}=
\hat{a}_q\exp{(i\delta_{q,j}^{(a)})}$, and similarly for the
output photon mode $\hat{b}_{q,j}=
\hat{b}_q\exp{(i\delta_{q,j}^{(b)})}$. The read-in and read-out
processes are assumed to require a time $t_p$ each. In general the
couplings $\Omega_{q,j}$ and $\tilde{\Omega}_{q,j}$ will depend on
time $t$. In the following we assume a simple time dependence
where the magnitude of the read-in (read-out) coupling is switched
on to a constant value for the time $t_p$ that maximizes storage
(retrieval), then switched off. For simplicity we also let
$|\Omega_{q,j}|= |\tilde{\Omega}_{q,j}|\forall j$ -- the
generalization to different couplings is straightforward.
Inhomogeneous broadening can furthermore lead to phases linearly
increasing with time, and during storage some additional dephasing
can occur. As a result of these assumptions we write
$\Omega_{q,j}=\kappa_{q,j}e^{\mathrm{i}(K_{q,j}t)}$ and
$\tilde{\Omega}_{q,j}=\kappa_{q,j}e^{\mathrm{i}[M_{q,j}t+f_{q,j}(t_s)]}$,
where $f_{q,j}(t_s)$ appears as a result of eliminating
$\hat{H}_{S,q}^{(j)}$ from the dynamics. The parameters
$\kappa_{q,j}$, $K_{q,j}$, $M_{q,j}$, $\delta_{q,j}^{(a)}$,
$\delta_{q,j}^{(b)}$ and $f_{q,j}(t_s)$ are all assumed to be real
normally-distributed stochastic variables with respect to the
storage medium. For instance $K_{q,j}$ is broadened around a mean
value $\bar{K}_q$ by a width $w_{K,q}$ and so on.

To obtain an analytical expression for $\mathcal{F}$, we first
note that if $K_{q,j}=\bar{K}_q\forall j$ the system reduces to a
two-level problem, and the evolution during read-in can be solved
exactly. To this end we rewrite the read-in Hamiltonian
\begin{equation}
\label{pertham}
\hat{H}_q^{(j)}=\kappa_{q,j}[e^{\mathrm{i}\bar{K}_qt}+e^{\mathrm{i}\bar{K}_qt}(e^{\mathrm{i}\delta_{q,j}^{(K)}t}-1)]\hat{a}_{q,j}\hat{\sigma}_{eg}^{(j)}+\mathrm{H.c.}\nonumber
\end{equation}
and treat the term containing the fluctuation $\delta_{q,j}^{(K)}$
in $K_q$ perturbatively up to second order, and similarly for
$\hat{\tilde{H}}_q^{(j)}$. Since any mean broadening could be
corrected for, we assume that $\bar{K}_q=\bar{M}_q=0$ for
simplicity. The general initial normalized photon and auxiliary
qubit state can be written as
$\ket{\phi_0}=\alpha\ket{\underline{0}0}+\beta\ket{\underline{0}1}+
\gamma\ket{\underline{1}0} +\eta\ket{\underline{1}1}$. For each
component of $\ket{\phi_0}$ the evolution operator $\hat{U}$
according to $\sum_j\hat{H}_q^{(j)}$ and
$\sum_j\hat{\tilde{H}}_q^{(j)}$ can be used to calculate the final
wavefunction of the system at time $t_f=t_s+2 t_p$. Averaging over
the ensemble similarly to \cite{DCZ} allows us to rewrite
Eq.~(\ref{charmem}) as
\begin{equation}
\label{average}
\Lambda_M\otimes\mathbb{I}:\outprod{\phi_0}{\phi_0}\rightarrow\left\langle\left\langle
\hat{U}(\outprod{\phi_0}{\phi_0}\otimes\outprod{G}{G})\hat{U}^{\dagger}\right\rangle\right\rangle\mbox{,}
\end{equation}
where $\left\langle\left\langle\dots\right\rangle\right\rangle$
denotes averaging over the stochastic Hamiltonian variables then
tracing out the memory. Since $\ket{\phi_0}$ is normalized $\cal F$
can be calculated by minimizing over a single parameter
$X=\left|\alpha\right|^2+\left|\beta\right|^2$ in Eq.~(\ref{fid1}).
This results in
\begin{eqnarray}
\label{fid3}
{\cal F}&=&\left\{X_0^2\big\langle\big\langle\left|b_0\right|^2\big\rangle\big\rangle+2X_0(1-X_0)\mathrm{Re}\left\{\big\langle\big\langle b_0b_1^*\big\rangle\big\rangle\right\}\right.\nonumber\\
&&\left.+(1-X_0)^2\big\langle\big\langle\left|b_1\right|^2\big\rangle\big\rangle\right\}\mbox{,}
\end{eqnarray}
where $X_0$ is the value of $X$ that achieves the minimization in
Eq.~(\ref{fid1}), and $b_q$ is the amplitude of the final output
photon in logical state $q$. Differentiating $\mathcal{F}$ with
respect to $X$ gives a minimum of
\begin{equation}
\label{minparam}
X_0=\frac{\big\langle\big\langle\left|b_1\right|^2\big\rangle\big\rangle-\mathrm{Re}\left\{\big\langle\big\langle
b_0b_1^*\big\rangle\big\rangle\right\}}
{\big\langle\big\langle\left|b_0\right|^2\big\rangle\big\rangle-2\mathrm{Re}\left\{\big\langle\big\langle
b_0b_1^*\big\rangle\big\rangle\right\}+\big\langle\big\langle\left|b_1\right|^2\big\rangle\big\rangle}\mbox{,}
\end{equation}
but if this value lies outside $[0,1]$ then $X_0=0$ or $1$.
Applying second-order perturbation theory to the read-in and
read-out processes as previously described gives
\begin{eqnarray}
\label{bsquared}
&&\big\langle\big\langle\left|b_q\right|^2\big\rangle\big\rangle=\bigg[1-\frac{\Theta(w_{K,q}^2+w_{M,q}^2)}{8N(\bar{\kappa}_q^2+w_{\kappa,q}^2)}\bigg]\bigg[\frac{1}{N}+\nonumber\\
&&\qquad+\frac{(N-1)}{N(1+\tilde{w}_{\kappa,q}^2)^2}e^{-(w_{a,q}^2+w_{b,q}^2+w_{f,q}(t_s)^2)}\bigg]\mbox{,}\\
&&\mathrm{Re}\left\{\big\langle\big\langle
b_0b_1^*\big\rangle\big\rangle\right\}=\prod_{q=0,1}\bigg[\frac{e^{-(w_{a,q}^2+w_{b,q}^2+w_{f,q}(t_s)^2)/2}}{1+\tilde{w}_{\kappa,q}^2}\bigg]\nonumber\\
&&\qquad\times\bigg[1-\sum_{q=0,1}\frac{(4+\pi^2)\Theta(w_{K,q}^2+w_{M,q}^2)}{64N^2(\bar{\kappa}_q^2+w_{\kappa,q}^2)}\bigg]\mbox{,}
\end{eqnarray}
where $\tilde{w}_{\kappa,q}=w_{\kappa,q}/\overline{\kappa}_q$, and
$\Theta=(1+6\tilde{w}_{\kappa,q}^2+3\tilde{w}_{\kappa,q}^4)/(1+\tilde{w}_{\kappa,q}^2)^2$.
We see that $\mathcal{F}$ decreases exponentially in $w_{a,q}$,
$w_{b,q}$ and $w_{f,q}(t_s)$, and also decreases as both
$\tilde{w}_{\kappa,q}$ and $w_{x,q}^2/\overline{\kappa_q^2}$
increase ($x=K,M$). Due to the factors of $1/N$ appearing in these
latter terms, it is the exponential terms that will dominate for
large $N$. Let us also note that to obtain maximum absorption and
emission we set $t_p=\pi/2(\overline{\kappa_q^2}N)^{1/2}$, so the
terms containing $w_{x,q}$ could alternatively be seen to depend
quadratically on $t_p$. Finally, we observe that sufficient
conditions for $\mathcal{F}\lesssim1$ are that
$w_{a,q},w_{b,q},w_{f,q}(t_s)\ll1$ and
$\omega_{K,q},\omega_{M,q}\ll\overline{\kappa^2}$, with the latter
becoming less important as $N\rightarrow\infty$.

The value of $X_0$ represents the class of states that achieve the
minimum required in Eq.~(\ref{fid1}). To illustrate this let us
consider some special cases. (i) If the states $\ket{\underline
0}$ and $\ket{\underline{1}}$ of the photon are absorbed and
emitted in the same way ($b_0=b_1$), then evaluating $X_0$ gives
an indeterminate answer, reflecting the fact that $\mathcal{F}$ is
minimized by several choices of $\ket{\phi_0}$. Evaluation of
$\mathcal{F}$ in this situation gives a value
$\mathcal{F}=\big\langle\big\langle\left|b_0\right|^2\big\rangle\big\rangle$.
(ii) If state $\ket{\underline{1}}$ is perfectly stored, but state
$\ket{\underline{0}}$ is not stored at all, then $b_{0}=0$,
$X_0=1$, and $\mathcal{F}=0$. We also compare our measure with the
previously-defined fidelity $F_A$. If entanglement is preserved
i.e.~${\cal F}=1$ then $F_A=1$ if and only if the output photon
has the same mode function as the input photon. In the case where
the photon is stored and emitted with $100\%$ probability, but
becomes completely decorrelated with the auxiliary qubit $F_A$
could vary between $0$ and $1$ depending on the spatial mode
function of the output photon, but ${\cal F}=1/2$.

\begin{figure}
\begin{center}
\includegraphics[width=5cm]{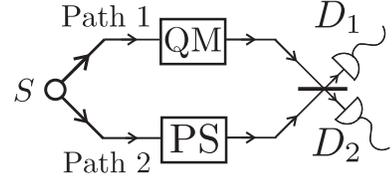}
\end{center}
\caption{Experimental method of measuring $\cal F$ requiring
storage of one logical state only. A source $S$ produces a
separable pair of photons, so that photon 1 is stored in the QM
and photon 2 enters the pulse shaper (PS). The photons interfere
at a beam splitter (the PS includes a time delay), and coincidence
measurements at detectors $D_1$ and $D_2$ are made.}
\label{Fig_expt}
\end{figure}

We now describe an experimental setup (shown in
Fig.~\ref{Fig_expt}) that, assuming case (i) above holds, would
allow us to measure $\mathcal{F}$. After read-out but before the
beam splitter (BS) the state of the photons will be
$\hat{\rho}_{\mathrm{in}}=\hat{a}_{PS}^{\dagger}\bigg\{\sum_m[p_m(\hat{b}_m^{\mathrm{out}})^{\dagger}\outprod{\mathrm{vac}}{\mathrm{vac}}\hat{b}_m^{\mathrm{out}}]+p_0\outprod{\mathrm{vac}}{\mathrm{vac}}\bigg\}\hat{a}_{PS}$,
where $\ket{\mathrm{vac}}$ denotes the vacuum, $\hat{a}_{PS}$ is
the annihilation operator for the mode of photon $2$ after the
pulse shaper (PS), and $\left\{\hat{b}_m^{\mathrm{out}}\right\}$
with $m\geq 1$ is the set of annihilation operators corresponding
to the eigenmodes of the state of photon $1$. The eigenvalues are
in descending order $p_1\geq p_2\geq\dots$ and $p_0$ is the
probability of not retrieving the photon on demand. Noting that
most detectors cannot resolve photon number, the probability of
obtaining a click in one of the detectors $D_j$ ($j=1,2$) is $
P_j=\sum_{m=1}^{\infty}p_m(1+{\cal O}_m)/4+p_0/2$, and of a
detection in both $D_1$ and $D_2$ is $
P_{12}=\sum_{m=1}^{\infty}p_m(1-{\cal O}_m)/2$, where ${\cal O}_m
= |\bra{\rm vac}\hat{a}_{PS} (\hat{b}_m^{\mathrm{out}}
)^\dagger\ket{\mathrm{vac}}|^2$ is the overlap of the field modes
after the BS. Both the minimum value of $P_{12}$ and the maximum
value of $P_1+P_2$ are obtained when $\hat{a}_{PS} =
\hat{b}_1^{\mathrm{out}}$ i.e.~when the mode of photon $2$ is
precisely the dominant mode of photon $1$ and for this setting
$p_0+p_1=P_1+P_2-P_{12}$. Hence by tuning the PS the dominant mode
of the memory photon can be found experimentally. This tuning then
corresponds to the $\hat{U}_M$ that maximizes $\mathcal{F}$ as in
Eq.~(\ref{fid1}). We can then deduce
$p_1=\big\langle\big\langle\left|b_0\right|^2\big\rangle\big\rangle$
by removing the beam splitter and measuring the probability of the
memory photon not being re-emitted on demand. Therefore
$\mathcal{F}$ can be deduced.

\begin{figure}
\begin{center}
\includegraphics[width=8.5cm]{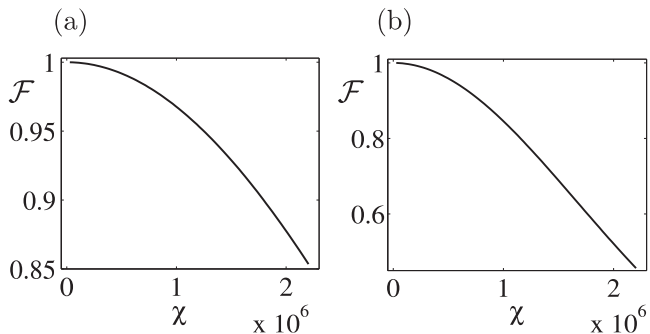}
\end{center}
\caption{The entanglement fidelity of a Raman QM with (a)
$\zeta=4.49\times10^{-14}$, (b) $\zeta=2.25\times10^{-13}$. In
both cases the photon bandwidth $\delta_p=0.1\Delta$, with
$\Delta=10^{13}\mathrm{s}^{-1}$. The atomic level splittings used
are
$\ket{g}\rightarrow\ket{\mathrm{int}}=5\times10^{15}\mathrm{s}^{-1}$
and
$\ket{e}\rightarrow\ket{\mathrm{int}}=3.5\times10^{15}\mathrm{s}^{-1}$,
and the ensemble consisted of $N=10^8$ atoms.} \label{Fig_temp}
\end{figure}

We conclude our analysis by applying the fidelity measure
$\mathcal{F}$ to a specific memory setup. We determine
$\mathcal{F}$ for a QM for one single photon state based on
off-resonant stimulated Raman scattering in an ensemble of
$\Lambda$-atoms \cite{nunn}. The atoms each have mass $M$ and
temperature $T$, and have the same internal level structure as the
general absorbers considered in Fig.~\ref{Fig_setup}(a). The
photon is incident on the ensemble and excites the
$\ket{g}\rightarrow\ket{\mathrm{int}}$ transition. A control field
drives $\ket{\mathrm{int}}\leftrightarrow\ket{e}$ and stores the
photon as a collective excitation in the ensemble. The probe and
control fields are assumed to co-propagate with carrier
wavevectors of magnitude $k_p$ and $k_c$ respectively. Retrieval
of the photon is achieved by applying another control field a time
$t_s$ after read-in. We assume that the probe and control fields
are both far-detuned (detuning $\Delta$) from level
$\ket{\mathrm{int}}$, so this state can be adiabatically
eliminated giving a medium consisting effectively of two-level
atoms. Therefore the main source of stochastic variation in the
coupling of the atoms to the photon arises from the atomic motion,
which we treat semiclassically assuming a Boltzmann distribution
for atomic velocity components $v_j$ in the direction of the field
propagation. This leads to $K_{q,j}=M_{q,j}=v_j\omega_c/c$,
$f_{q,j}(t_s)=\chi v_j/c$ and widths given by
$w_{K,q}=w_{M,q}=\omega_c\zeta^{1/2}$, $w_{f,q}=\chi\zeta^{1/2}$,
where $c$ is the speed of light, $\zeta=k_BT/Mc^2$,
$\chi=(k_p-k_c)ct_s$, and $\omega_c=c k_c$. In this scheme
$f_{q,j}(t_s)$ originates from the motion of the atoms during the
storage time and  $K_{q,j}$ and $M_{q,j}$ arise from the
Doppler-shifting of the field frequencies. The parameter
$\kappa_q=\kappa$ is defined by the couplings of the photon and
control fields to the atoms, and is assumed to be a constant.

The amplitude of the final photon state can be calculated as for
the general case, which upon substitution into Eq.~(\ref{fid3})
yields the following expression for $\mathcal{F}$ up to $N^{-2}$,
\begin{equation}
\label{ramanbsquared}
\mathcal{F}=e^{-\chi^2\zeta/2}(1-\chi^2\zeta)\bigg[e^{-\chi^2\zeta/2}(1-\chi^2\zeta)-\frac{3\zeta\omega_c^2}{2\kappa^2N^2}\bigg].
\end{equation}
We see that the two main contributions to the decrease in
$\mathcal{F}$ are the Doppler broadening terms, which are
quadratic in the ratio $\omega_c\sqrt{\zeta}/\kappa$, and the
storage time dephasing terms, which depend on $\chi^2\zeta$. This
observation results in the requirement that $\chi\sqrt{\zeta}\ll
1$ in order to achieve $\mathcal{F}\lesssim1$ for $N\gg1$.
Fig.~\ref{Fig_temp} shows the entanglement fidelity of the Raman
quantum memory for two different values of $\zeta$, and the
expected decrease in $\mathcal{F}$ with increasing $\chi$ is
observed.

In summary we have introduced a figure of merit $\mathcal{F}$ for
a general QM based on gate fidelity and derived an analytical
expression for it. Our calculations took into account stochastic
fluctuations in the coupling parameters whose origin might vary
for different QM schemes. We concluded by applying our formalism
to a specific atomic quantum memory.

\begin{acknowledgments}
This work was supported by the EPSRC (UK) through the QIP IRC
(GR/S82716/01) and project EP/C51933/01. JN thanks Hewlett-Packard
and FCW thanks Toshiba for support. DJ acknowledges discussions
with N.~L\"{u}tkenhaus, R.~Renner and D.~Bru{\ss}. The research of
DJ was supported in part by The Perimeter Institute for
Theoretical Physics. IAW was supported in part by the European
Commission under the Integrated Project Qubit Applications (QAP)
funded by the IST directorate as Contract Number 015848.
\end{acknowledgments}

\end{document}